\documentclass[twocolumn]{aastex63}

\usepackage{color}
\usepackage{upgreek}

\usepackage{threeparttable}

\newcommand{\p}{\partial}

\newcommand\cqg{{Class. Quantum Grav.}}

\def\sun{_\odot}

\def\Dwa{$\,$\uppercase\expandafter{\romannumeral5}$\,$}

\def\sless{\lower2pt\hbox{$\buildrel {\scriptstyle <}
   \over {\scriptstyle\sim}$}}

\def\sgreat{\lower2pt\hbox{$\buildrel {\scriptstyle >}
   \over {\scriptstyle\sim}$}}
\def\sharpnull#1{}

\received{\today}
\revised{\today}
\accepted{\today}
\shorttitle{A magnetar engine for short GRBs and kilonovae}
\shortauthors{M\"osta al.}
\graphicspath{{./}{figures/}}

\begin{document}

\title{A magnetar engine for short GRBs and kilonovae}

\correspondingauthor{Philipp M\"osta}
\email{p.moesta@uva.nl}

\author[0000-0002-0786-7307]{Philipp M\"osta} 
\affil{GRAPPA, Anton Pannekoek Institute for Astronomy and Institute of
High-Energy Physics, University of Amsterdam, Science Park 904, 1098 XH
Amsterdam, The Netherlands}

\affil{Department of Astronomy,
University of California at Berkeley, 501 Campbell Hall, Berkeley, CA 94720,
USA}

\author[0000-0001-6982-1008]{David Radice} 
\affil{Institute for Gravitation \& the Cosmos, The Pennsylvania State University, University Park, PA 16802, USA}
\affil{Department of Physics, The Pennsylvania State University, University Park, PA 16802, USA}
\affil{Department of Astronomy \& Astrophysics, The Pennsylvania State University, University Park, PA 16802, USA}

\author[0000-0003-1424-6178]{Roland Haas} 
\affil{NCSA, University of Illinois, Urbana-Champaign, USA}

\author[0000-0002-4518-9017]{Erik Schnetter} 
\affil{Perimeter Institute for Theoretical Physics, Waterloo, ON, Canada}
\affil{Department of Physics and Astronomy, University of Waterloo, Waterloo, Ontario, Canada}
\affil{Center for Computation \& Technology, Louisiana State University, Baton Rouge, USA}

\author[0000-0002-2334-0935]{Sebastiano Bernuzzi} 
\affil{Theoretisch-Physikalisches Institut, Friedrich-Schiller-Universit\"at Jena, 07743,
Jena, Germany}



\begin{abstract}
	We investigate the influence of magnetic fields on the evolution of
	binary neutron-star (BNS) merger remnants via three-dimensional (3D)
	dynamical-spacetime general-relativistic (GR) magnetohydrodynamic (MHD)
	simulations. We evolve a postmerger remnant with an initial poloidal
	magnetic field, resolve the magnetoturbulence driven by shear flows,
	and include a microphysical finite-temperature equation of state (EOS).
	A neutrino leakage scheme that captures the overall energetics and
	lepton number exchange is also included. We find that turbulence
	induced by the magnetorotational instability (MRI) in the hypermassive
	neutron star (HMNS) amplifies magnetic field to beyond
	magnetar-strength ($10^{15}\, \mathrm{G}$).  The ultra-strong toroidal
	field is able to launch a relativistic jet from the HMNS. We also find
	a magnetized wind that ejects neutron-rich material with a rate of
	$\dot{M}_{\mathrm{ej}} \simeq 1 \times10^{-1}\, \mathrm{M_{\odot}\,
	s^{-1}}$. The total ejecta mass in our simulation is $5\times 10^{-3}\,
	\mathrm{M_{\odot}}$. This makes the ejecta from the HMNS an important
	component in BNS mergers and a promising source of $r$-process elements
	that can power a kilonova. The jet from the HMNS reaches a terminal
	Lorentz factor of $\sim 5$ in our highest-resolution simulation. The
	formation of this jet is aided by neutrino-cooling preventing the
	accretion disk from protruding into the polar region. As neutrino
	pair-annihilation and radiative processes in the jet (which were not
	included in the simulations) will boost the Lorentz factor in the jet
	further, our simulations demonstrate that magnetars formed in BNS
	mergers are a viable engine for short gamma-ray bursts (sGRBs). 
\end{abstract}



\section{Introduction} \label{sec:intro}

The inspiral and merger of two neutron star (NSs) are among the loudest and
most luminous events in the universe~\citep{ligo:17nsns,ligo:17nsnsmm}.
Radioactive material ejected during and after the merger powers a kilonova
transient and creates the heaviest elements in the universe~\citep{kasen:17}.
Jetted outflows from the merger remnant can launch a
sGRB~\citep{ruiz:16,goldstein:17,savchenko:17}. The multimessenger observations
of GW170817 have confirmed our basic understanding of NS mergers
(NSMs)~\citep{metzger:17} but two key open astrophysics problems for NSMs are
how to generate fast-enough outflows to explain the observed blue kilonova
component in GW170817 and whether magnetars can launch sGRB
jets~\citep{dai:98a,zhang:01}.

Follow-up of late-time kilonova emission and sGRB radio
observations~\citep{mooley:18a,ghirlanda:19} have begun to constrain different
engine models but no conclusion on the nature of the engine for GRB170817 (i.e.
black-hole or magnetar) has been reached. ~\citet{metzger:18a}
have suggested a magnetar-origin for the blue kilonovae component because
hydrodynamic simulations have not been able to produce fast enough
outflows~\citep{fahlman:18}. Similarly, ~\citet{bucciantini:12} have shown
that magnetars left behind by a NSM are capable of explaining sGRBs, but
Numerical Relativity simulations of NSMs have only been able to produce jets
after black-hole formation~\citep{ruiz:16}. Simulations leaving behind a stable
magnetar have found that baryon pollution of the polar region prevents the
launch of a sGRB jet~\citep{ciolfi:17,ciolfi:19,ciolfi:20b}, but these
simulations did not include neutrino effects.


Any merger remnant is likely magnetized by seed fields of the individual NSs
and their amplification via Kelvin-Helmholtz instability in the shear layer
during the merger~\citep{obergaulinger:10,zrake:13,kiuchi:15}. As a result,
magnetic fields play a key role in the postmerger evolution of HMNS. They can
launch outflows that eject material along the rotation axis of the
remnant~\citep{kiuchi:12, siegel:14} and remove mass and angular momentum.
Inside the remnant and in the accretion disk magnetoturbulence can act to
redistribute angular momentum and launch winds from the disk surface. 

There has been substantial previous work modeling NSMs via MHD
simulations,
e.g.~\citet{price:06,duez:06b,anderson:08b,rezzolla:11,giacomazzo:11,dionysopoulou:13,neilsen:14,palenzuela:15,ruiz:16,ciolfi:19,ruiz:19,ciolfi:20b},
but these simulations did not employ high-enough resolution to capture the
turbulent magnetic field evolution in the merger remnant. Notable exceptions
are \citet{kiuchi:15,kiuchi:17} which have performed the highest-resolution
GRMHD simulations of NSMs and post-merger evolution to date, but these
simulations did not include a realistic nuclear EOS or neutrinos.

\begin{figure*}[t] 
\centering
	\includegraphics[width=0.32\textwidth]{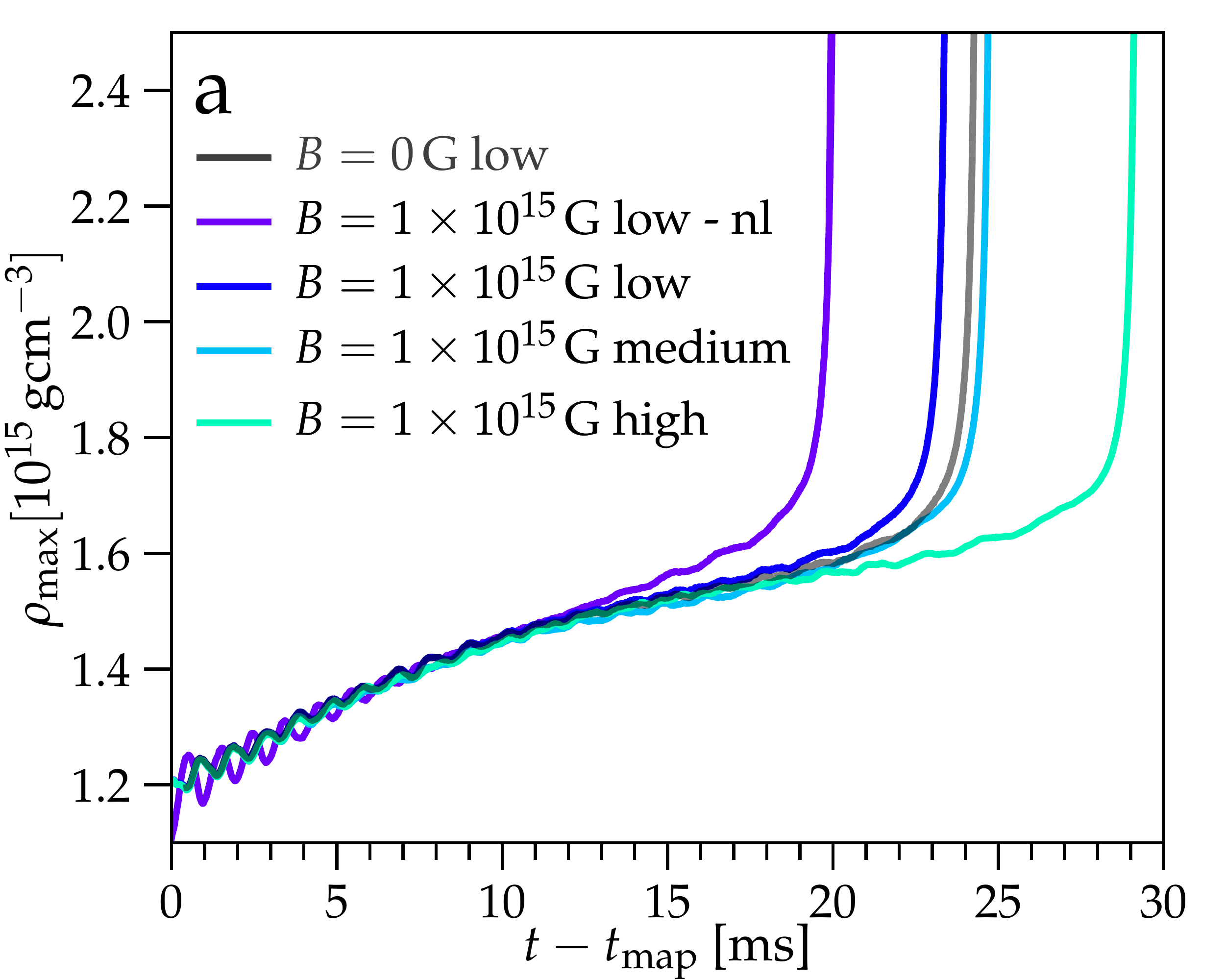}
	\includegraphics[width=0.32\textwidth]{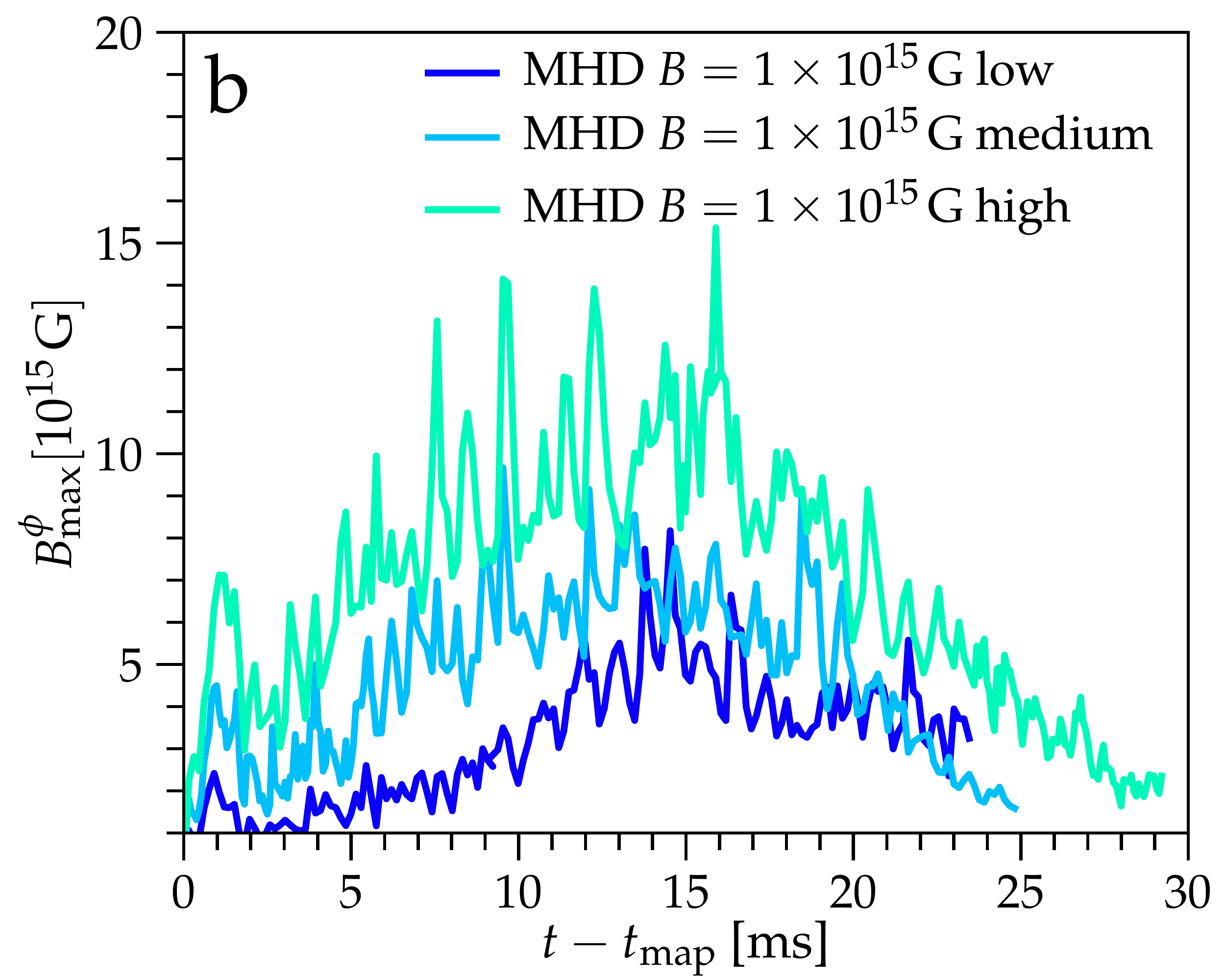}
	\includegraphics[width=0.32\textwidth]{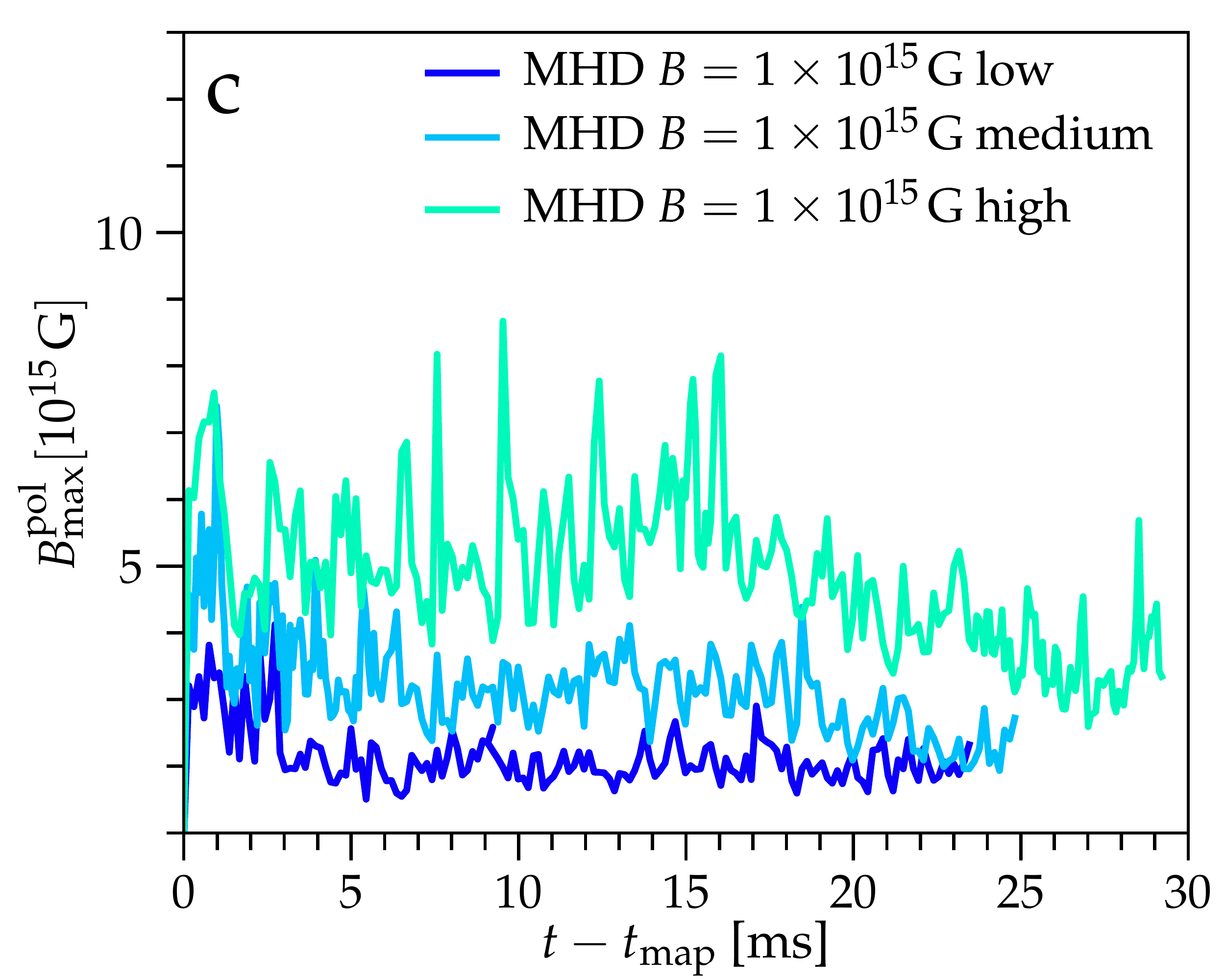}
	\vspace{0.1cm} 
	\caption{(a) Maximum density as a function of
	post-mapping time $t-t_{\mathrm{map}}$ for simulations B0 (black),
	B15-nl (magenta), B15-low (blue), B15-med (cyan), and B15-high (light
	green).  (b) Maximum toroidal magnetic field strength as a function of
	post-mapping time $t-t_{\mathrm{map}}$ for simulations B15-low (blue),
	B15-med (cyan), and B15-high (light green).  (c) Maximum poloidal
	magnetic field strength as a function of post-mapping time
	$t-t_{\mathrm{map}}$ for simulations B15-low (blue), B15-med (cyan),
	and B15-high (light green).} 
\label{fig:rhobmax} 
\vspace{0.5cm} 
\end{figure*}
We perform high-resolution dynamical-spacetime GRMHD simulations of NS merger
remnants including a nuclear EOS and neutrino effects. For comparison we 
also perform simulations which either do not include a magnetic
field or do not include neutrino effects. We initialize our
simulations by mapping a BNS merger simulation performed in GRHD to a
high-resolution domain and add a poloidal magnetic field. We find that
MRI-induced turbulence in the HMNS amplifies the magnetic field to beyond
magnetar-strength in the HMNS. The added and amplified field launches a
relativistic jet from the HMNS in the simulations that include neutrino
effects. The emergence of this jet is aided by neutrino cooling reducing
baryon pollution in the polar region compared to simulations without neutrino
effects. The jet reaches a terminal Lorentz factor of $\sim 5$ in our
highest-resolution simulation. In all simulations a magnetized wind driven from
the HMNS~\citep{thompson:04b} ejects neutron-rich material with a rate of
$\dot{M}_{\mathrm{ej}} \simeq 1 \times10^{-1}\, \mathrm{M_{\odot}\, s^{-1}}$
accounts for the majority of ejected material. The total ejecta mass is
$5\times 10^{-3}\, \mathrm{M_{\odot}}$. This makes the ejecta from the HMNS an
important component in BNS in addition to the dynamical ejecta and the disk
wind once a BH has formed. Our simulations demonstrate that neutrino effects
can prevent baryon pollution of the polar region in NSM remnants and that
magnetars formed in NS mergers are a viable engine for both sGRBs and kilonova
if a large-scale dipolar field can be created.

This paper is organized as follows. In Sec.~\ref{sec:methods}, we present the
physical and computational setup and numerical methods used. In
Sec.~\ref{sec:dynamics}, we present the simulation dynamics, followed by a 
description of the jet and ejecta dynamics in Sec.~\ref{sec:outflows}. 
We conclude with a discussion of our findings in Sec.~\ref{sec:discussion}.

\section{Numerical Methods and Setup} \label{sec:methods}
 
We employ ideal GRMHD with adaptive mesh refinement (AMR) and spacetime
evolution provided by the open-source \texttt{Einstein Toolkit}
~\citep{EinsteinToolkit:2019_10,et:12,	Schnetter:2003rb,Goodale:2002a} module
\texttt{GRHydro}~\citep{moesta:14a}. GRMHD is implemented in a finite-volume
fashion with WENO5 reconstruction~\citep{reisswig:13a, tchekhovskoy:07} and the
HLLE Riemann solver \citep{HLLE:88} and constrained transport \citep{toth:00}
for maintaining $\mathrm{div} \vec{B} = 0$. We employ the $K_0 =
220\,\mathrm{MeV}$ variant of the equation of state of \cite{lseos:91} and the
neutrino leakage/heating approximations described in \cite{oconnor:10} and
\cite{ott:13a}. The scheme tracks three species: electron neutrinos $\nu_e$, 
electron antineutrinos $\bar{\nu_e}$, and heavy-lepton neutrinos which are grouped 
together into a single species $\nu_x$. The scheme approximates neutrino
cooling by first computing the energy-averaged neutrino optical depths along radial
rays and in a second step calculating local estimates of energy and lepton loss
rates. We employ 20 rays in $\theta$, covering $[0,\pi/2]$,and 40 rays in $\phi$
covering $[0, 2\pi]$. Each ray has 800 equidistant points to ~120 km and 200
logarithmically spaced points covering the remainder of the ray. Neutrino
heating is approximated using the neutrino heating rate
$$\mathcal{Q}_{\nu_i}^{\mathrm{heat}} = f_{\mathrm{heat}}
\frac{L_{\nu_i}(r)}{4\pi r^2} S_{\nu}
\left\langle\epsilon_{\nu_i}^2\right\rangle \frac{\rho}{m_n} X_i
\left\langle\frac{1}{F_{\nu_i}}\right\rangle e^{-2\tau_{\nu_i}},$$ where
$L_{\nu_i}$ is the neutrino luminosity emerging from below as predicted by the
neutrino leakage approximation along radial rays, $S_{\nu}=0.25
(1+3\alpha^2)\sigma_0(m_e c^2)^{-2}$, $\sigma_0 = 1.76 \times 10^{-44}\,
\mathrm{cm^2}$, $\alpha = 1.23$, $m_e$ the electron mass, $m_n$ the neutron
mass, $c$ the speed of light, $\rho$ the rest-mass density, $X_i$ the neutron
(proton) mass fraction for electron neutrinos (antineutrinos),
$\left\langle\epsilon_{\nu_i}^2\right\rangle$ the mean-squared energy of the
$\nu_i$ neutrinos, and $\left\langle F_{\nu_i}^{-1}\right\rangle$ the mean
inverse flux factor.  $f_{\mathrm{heat}}$ , the heating scale factor, is a free
parameter in this scheme and we set $f_{\mathrm{heat}} = 1.05$, consistent with
heating in comparison to full neutrino transport schemes in core-collapse
supernova simulations~\citep{ott:13a}. Further details of the implementation of
the scheme can be found in \cite{oconnor:10, ott:12a}. We turn neutrino heating
off below a density of $\rho = 6.18\times 10^{10}\, \mathrm{g\, cm^{-3}}$ for
numerical stability. The leakage scheme employed here captures the overall
neutrino energetics correctly up to a factor of a few when compared to full
transport schemes in core-collapse supernova simulations~\citep{oconnor:10}.
It does not account for momentum deposition, energy dependence, or neutrino
pair annihilation. 
While the detailed composition of the ejecta depends sensitively on the
neutrino scheme we expect the main result of this study, the emergence of
a relativistic outflow, to hold. 

We map initial data from a GRHD BNS simulation performed with
\texttt{WhiskyTHC}, particularly model LS135135M0, an equal-mass binary
with individual neutron star masses at infinity $M_a = M_b = 1.35\, M_\odot$
and resolution $h \simeq 185\, \mathrm{m}$ covering the merger remnant from
\cite{radice:18}. The \texttt{WhiskyTHC} simulation uses the same EOS (LS220)
and a very similar but not identical implementation of the neutrino Leakage
approximation used in the simulations presented here. We map the HD simulation
at $t_{\mathrm{map}}-t_{\mathrm{merger}} = 17\,\mathrm{ms}$ and add a magnetic
field. The mapping time is chosen to avoid transient effects created by the
oscillatory behavior of the remnant core in the early postmerger evolution. We
set up the initial magnetic field using a vector potential of the form $$A_r =
A_\theta = 0; A_\phi = B_0 ({r_0^3})({r^3+r_0^3})^{-1}\, r \sin \theta,$$ where
$B_0$ controls the strength of the field. 
We choose $r_0 = 20\, \mathrm{km}$ to keep the field nearly constant inside the
HMNS. We choose to map this parameterized magnetic field to have full control
over our ability to resolve the MRI in the remnant. 
In doing so we implicitly assume the presence of a dynamo process producing a
large scale ordered magnetic field following field amplification during merger.
We caution the reader that, while the presence of such dynamo is plausible
\citep{moesta:15, raynaud:20}, current simulations do not have sufficient
resolution to resolve it \citep{zrake:13,kiuchi:15,kiuchi:17}.

We perform simulations for initial magnetic field strength $B_{0} = 10^{15}\,
\mathrm{G}$ (B15-nl, B15-low, B15-med, and B15-high) and a simulation with
$B_{0} = 0\, \mathrm{G}$ (B0) which acts as a hydrodynamic reference simulation
but is performed using the MHD code to keep the numerical methods identical between
the simulations. To investigate the influence of neutrino physics we also
perform a simulation with magnetic field $B_{0} = 10^{15}\, \mathrm{G}$ but
without the neutrino leakage scheme enabled (B15-nl). Simulations B0 and B15-nl 
are performed at the same resolution as B15-low.

\begin{figure}[t]
	\includegraphics[width=0.49\textwidth]{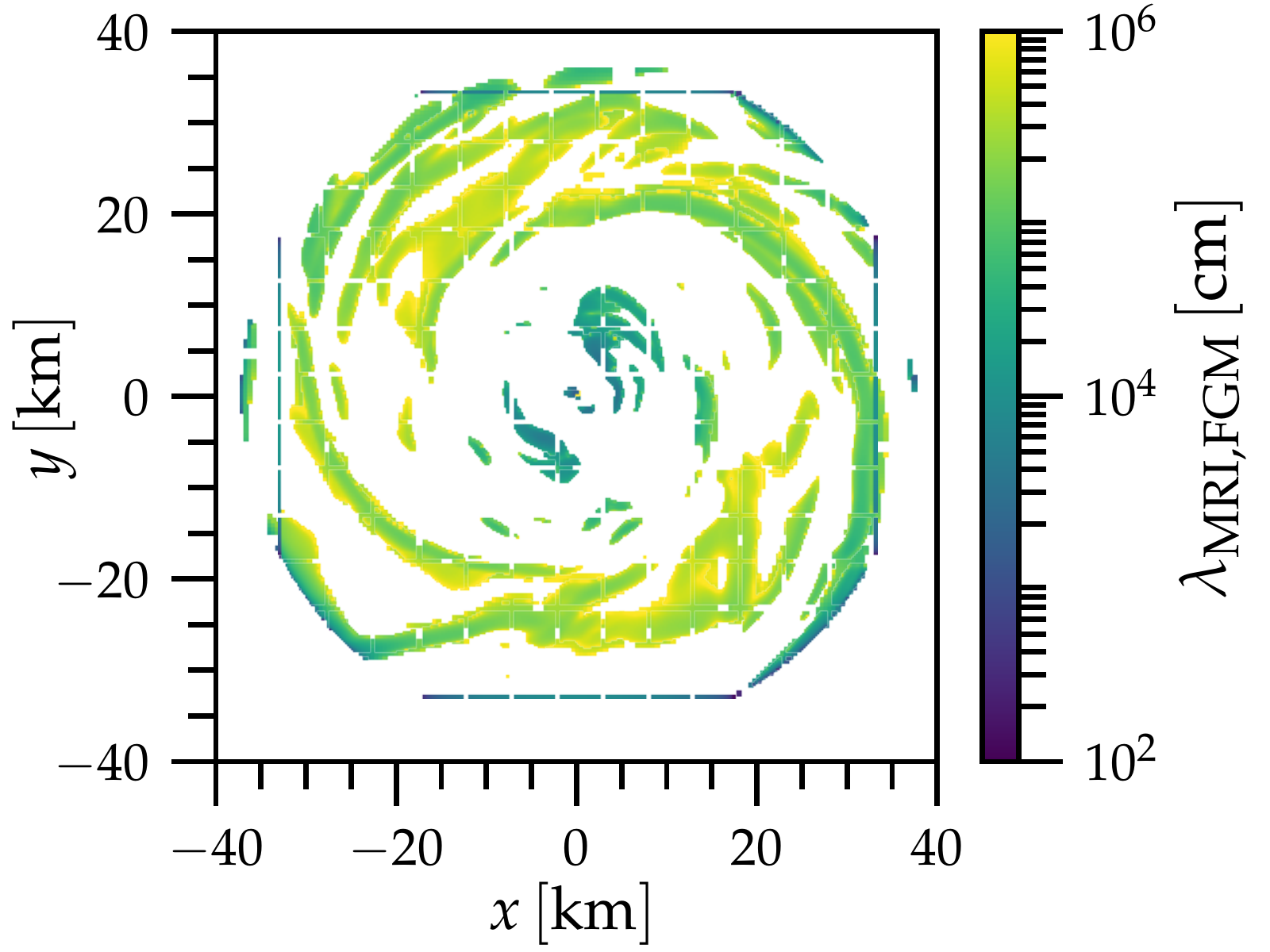}
\vspace{0.1cm}
	\caption{Equatorial slice ($xy$-plane at $z=1.4776\, \mathrm{km}$) of
	the wavelength of the fastest growing mode of the MRI zoomed in to show
	the innermost [-40km, 40km] for simulations B15-low at
	time $t-t_{\mathrm{map}} = 0\, \mathrm{ms}$.} 
\label{fig:2dlmri} 
\vspace{0.5cm} 
\end{figure}

\begin{figure*}[t]
	\includegraphics[width=0.99\textwidth]{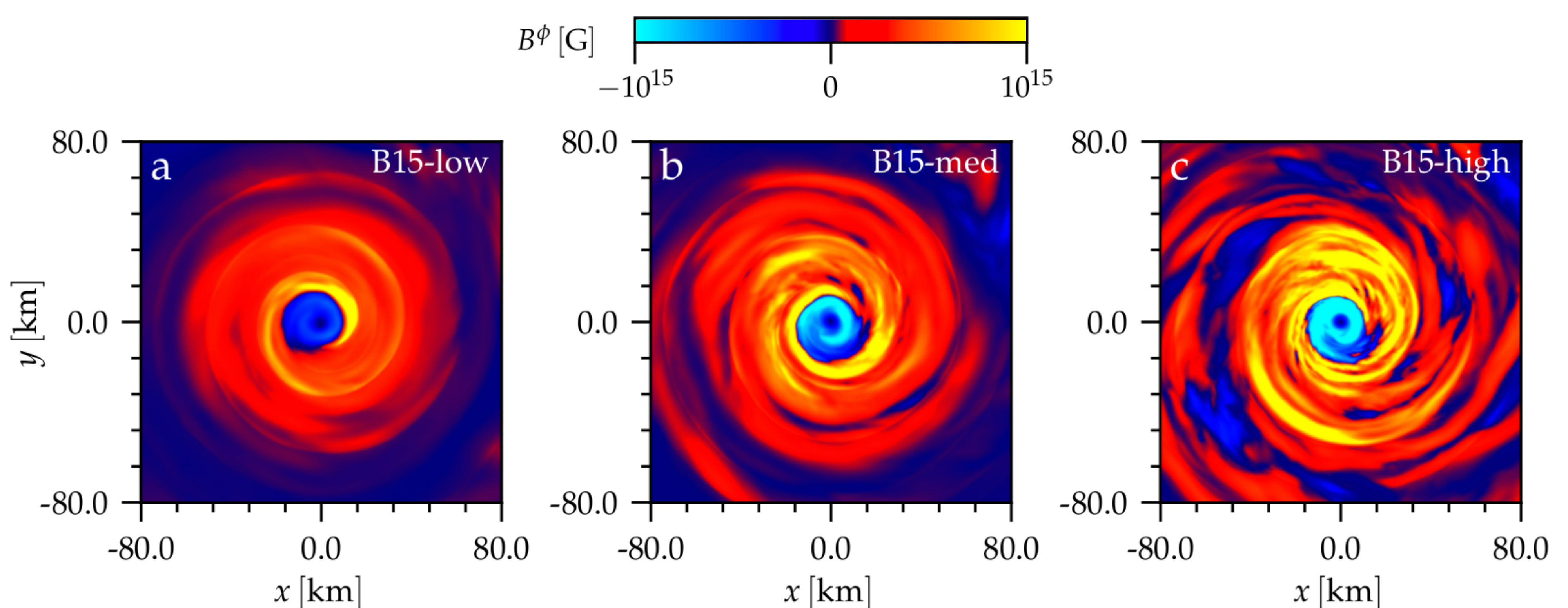}
\vspace{0.1cm}
	\caption{Equatorial slices ($xy$-plane at $z=1.4776\, \mathrm{km}$) of
	toroidal magnetic field strength $B^{\phi}$ zoomed in to show the
	innermost [-80km, 80km] for simulations B15-low (panel a),
	B15-med (panel b), and B15-high (panel c). All panels show the
	simulations at time $t-t_{\mathrm{map}} = 20.9\, \mathrm{ms}$.} 
\label{fig:2dbphi} 
\vspace{0.5cm} 
\end{figure*}

We use a domain with outer boundaries $\sim$$355\,\mathrm{km}$ and five AMR
levels in a Cartesian grid. The AMR grid structure consists of boxes with
extent [$177.3\,\mathrm{km}$, $118.2\,\mathrm{km}$, $59.1\,\mathrm{km}$,
$29.6\,\mathrm{km}$]. Refined meshes differ in resolution by factors of 2. We
perform simulations at three different resolutions. For our fiducial (low
resolution) simulations, the coarsest resolution is $h_{\mathrm{coarse}} =
3.55\, \mathrm{km}$ and the level covering the HMNS has $h_{\mathrm{fine}} =
220\, \mathrm{m}$. For our medium and high-resolution simulations we use
$h_{\mathrm{coarse}} = 1.77\, \mathrm{km}$ and $h_{\mathrm{fine}} = 110\,
\mathrm{m}$, and $h_{\mathrm{coarse}} = 0.89\, \mathrm{km}$ and
$h_{\mathrm{fine}} = 55\, \mathrm{m}$. 

We perform simulations in 3D with reflection symmetry in $z$-direction. To
prevent numerically-driven oscillations in the magnetic field, we apply
diffusivity and hyperdiffusivity at the level of the induction equation for the
magnetic field via a modified Ohm’s law. We choose $\vec{E} = - \vec{v} \times
\vec{B} + \eta \vec{J} - \eta_3 \nabla^3 \times \vec{B}$, where $\vec{J} =
\vec{\nabla} \times \vec{B}$ is the 3-current density. In this way the modified
Ohm's law does not impact the ability of the constrained transport scheme to
maintain the $\nabla \cdot \vec{B} = 0$ constraint.  
$\Delta x^i\, ||\nabla \cdot
\vec{B}||_2 / ||\vec{B}||_2 $ in our simulations is $\simeq 6 \times 10^{-8}$,
$\simeq 8 \times 10^{-9}$, and $\simeq 1 \times 10^{-9}$ for B15-low, B15-med,
and B15-high.  We set $\eta = 1.0 \times 10^{-2}$, $\eta = 5.0 \times 10^{-3}$,
and $\eta = 2.5 \times 10^{-3}$ for B15-low, B15-med, and B15-high, and $\eta_3
= 3.75 \times 10^{-3}$. We estimate the impact of the added diffusivity and
hyperdiffusivity terms by studying the time evolution of perturbations of the
magnetic field of the form $B_k(t) = B_k (t=0)\, e^{-k^2 \eta t}$ for $\eta$
and $B_k(t) = B_k (t=0)\, e^{-k^4 \eta_3 t}$ for $\eta_3$.  The condition for
the diffusivity term not to interfere with numerically resolving the fastest
growing mode (FGM) of the MRI can be expressed as $k^2 \eta <<
\frac{1}{\tau_{\mathrm{FGM,MRI}}}$. Using $\tau_{\mathrm{FGM,MRI}} \simeq 0.5\,
\mathrm{ms}$, $k = \frac{1}{20h}$ for $\lambda_{\mathrm{FGM,MRI}} \simeq 1000\,
\mathrm{m}$ (see Fig.~\ref{fig:2dlmri}), $h = 50\, \mathrm{m}$ in B15-high, and
expressing $\eta = \alpha h$ in terms of the grid spacing $h$ we can write this
condition as $\frac{\alpha}{h} << 4$. For B15-high with $\eta = 2.5 \times
10^{-3}$ we have $\frac{\alpha}{h} \simeq 0.07$. Following the same procedure
we find for the hyperdiffusivity parameter $k^4 \eta_3 <<
\frac{1}{\tau_{\mathrm{FGM,MRI}}}$ and $\frac{\beta}{h^3} << 1600$ with $\eta_3
= \beta h$. For $\eta_3 = 3.75 \times 10^{-3}$ in simulation B15-high we have
$\frac{\beta}{h^3} \simeq 18$. Thus the diffusivity and hyperdiffusivity terms
in our simulations operate on lengthscales significantly smaller than the
wavelength of the FGM of the MRI. (Hyper)diffusivity schemes are often employed
in high-order numerical simulations of magnetohydrodynamic turbulence,
e.g.~\citet{brandenburg:02}.

Material with density $\rho \leq 10^{4}\, \mathrm{g}\, \mathrm{cm}^{-3}$
in our simulations is considered part of the atmosphere and we set $v^i = 0$.

\section{Results} \label{sec:results}

\subsection{Overall dynamics and magnetic field evolution}
\label{sec:dynamics}

After mapping from the HD merger simulations to the postmerger MHD simulation
domain the added magnetic field in simulations B15-low, B15-med, and B15-high
adjusts over a few dynamical times ($t_{\mathrm{dyn,HMNS}} \simeq 0.5\,
\mathrm{ms}$ to the underlying hydrodynamical configuration of the remnant and
its accretion torus. There is amplification of both poloidal and toroidal
magnetic field within the first three milliseconds. A magnetized outflow forms
~\citep{kiuchi:12,siegel:14} and hoop stresses from the windup of strong
toroidal field along the rotation axis of the HMNS collimate part of this
outflow into a jet. This collimation does not appear in simulation B15-nl and
in simulation B0 only a neutrino-driven wind forms. The outflows persist until
the HMNS eventually collapses to a BH in all simulations. 

Fig.~\ref{fig:rhobmax} summarizes the overall dynamics of key quantities of the
HMNS evolution for simulations B0, B15-nl, B15-low, B15-med, and B15-high. Panel a)
shows the central density as a function of time after mapping
$t-t_{\mathrm{map}}$. The central density slowly increases as a function of
time for all simulations before the HMNS collapses to a BH. BH
formation occurs for simulation B15-nl after $\sim 19\, \mathrm{ms}$
and for simulation B0 after $\sim 23\, \mathrm{ms}$. Simulation
B15-low collapses $\sim 1\, \mathrm{ms}$ earlier than B0. Simulation
B15-med collapses to a BH $\sim 0.5\, \mathrm{ms}$ later than
simulation B15-low and B15-high collapses $\sim 6\, \mathrm{ms}$ later.

In Panels b and c we show the maximum toroidal and poloidal magnetic field
strength as a function $t-t_{\mathrm{map}}$ for simulations B15-low, B15-med,
and B15-high. After an initial nearly-instantaneous adjustment of the magnetic
field strength to the hydrodynamic flow, toroidal magnetic field is amplified
in all simulations. This growth saturates quickly for simulations B15-low and
B15-med but simulation B15-high, which fully resolves the fastest-growing mode
of the MRI, reaches a maximum toroidal field of $7\times 10^{15}\, \mathrm{G}$.  
The amplification happens predominantly in the shear
region outside the innermost core the HMNS (see Fig. 3 panel c). In
this region the FGM of the MRI has typical wavelengths of 500m - 2000m as shown
in Fig.~\ref{fig:2dlmri}. Our highest-resolution simulation B15-high covers this
wavelength with 10-40 points. The growth timescale (e-folding time) of
$\sim  0.5\, \mathrm{ms}$ approximately matches the rotation period of the
HMNS.  Subsequently, there is additional amplification of toroidal magnetic in
all simulations before the toroidal magnetic field strength decreases after
$t-t_{\mathrm{map}} \simeq 15\, \mathrm{ms}$.  The poloidal magnetic field is
similarly amplified within the first $\sim 2\, \mathrm{ms}$ but subsequently
remains in a turbulent state without additional amplification before decreasing
slightly in the last few ms before collapse to a BH. In the fully
turbulent state secondary instabilities and non-linear effects play an
important role and to capture these effects correctly much higher numerical
resolution than employed here is needed. For long-term fully sustained
turbulence physically complex and numerically difficult to resolve dynamo
processes are important. We do not see evidence for these in the
simulations presented here indicating that we are not fully resolving the
saturated turbulent evolution.

Fig.~\ref{fig:2dbphi} shows the toroidal magnetic field $B^{\phi}$ in the
$xy$-plane at $z=1.4776\, \mathrm{km}$ for simulations B15-low (panel a),
B15-med (panel b), and B15-high (panel c) a few ms before collapse to a
BH at $t-t_{\mathrm{map}} = 20.9\, \mathrm{ms}$. The colormap is chosen
such that yellow and light blue indicates magnetar-strength (or stronger)
toroidal magnetic field. For simulation B15-low in panel a only a single
cylindrical flow region outside the HMNS inner core with magnetar-strength
field is visible and barely any small-scale features are present. 
For simulation B15-med in panel b more magnetar-strength field is visible
and small-scale features start to emerge in the region of strong shear
outside the inner core of the HMNS $10\, \mathrm{km} < \omega < 40\,
\mathrm{km}$. For simulation B15-high in panel c the entire inner core and
shear region reach magnetar-strength field and small-scale features driven
by the magnetorotational turbulence are clearly visible and extend throughout
the entire shear region. We note that the inner region of negative toroidal
field in all simulations is a result of the positive angular velocity gradient
in the inner core.

\begin{figure*}[t]
\centering
	\begin{minipage}{0.25\textwidth}
	\vspace{-1.54cm}
	\includegraphics[width=\textwidth]{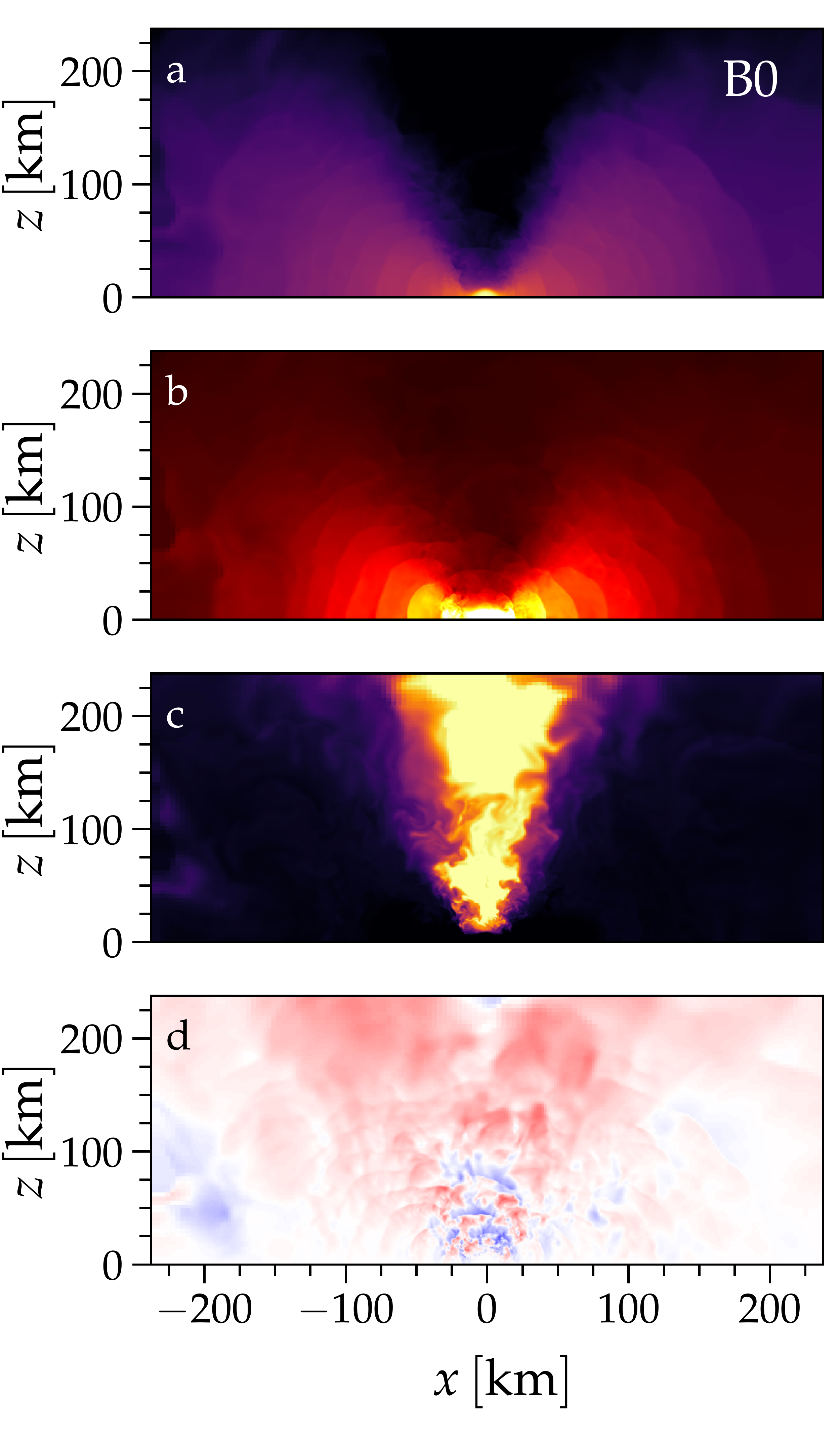}
	\end{minipage}
	\begin{minipage}{0.242\textwidth}
	\hspace{-0.4cm}
	\includegraphics[width=\textwidth]{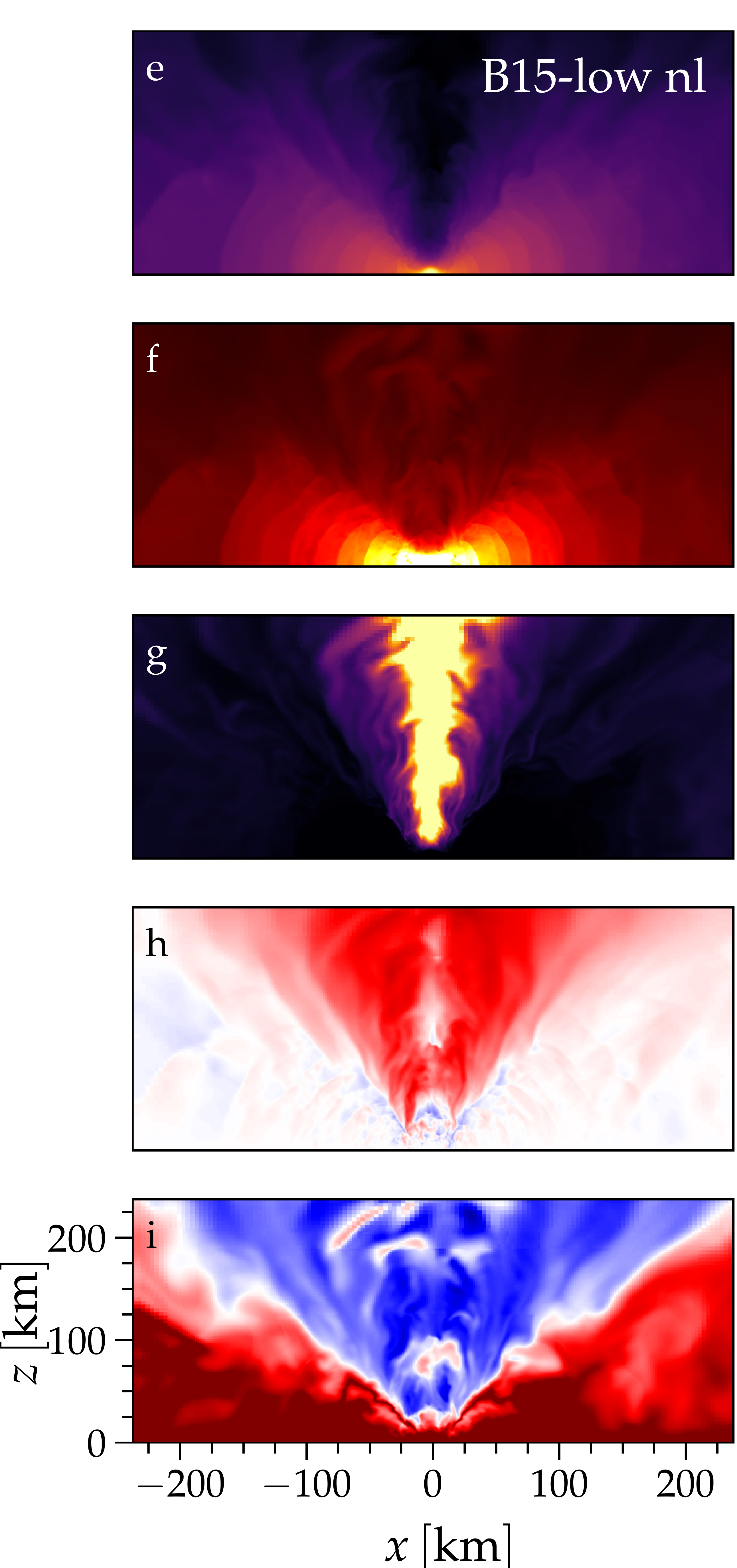} 
	\end{minipage}
	\begin{minipage}{0.206\textwidth}
	\hspace{-0.2cm}
	\includegraphics[width=\textwidth]{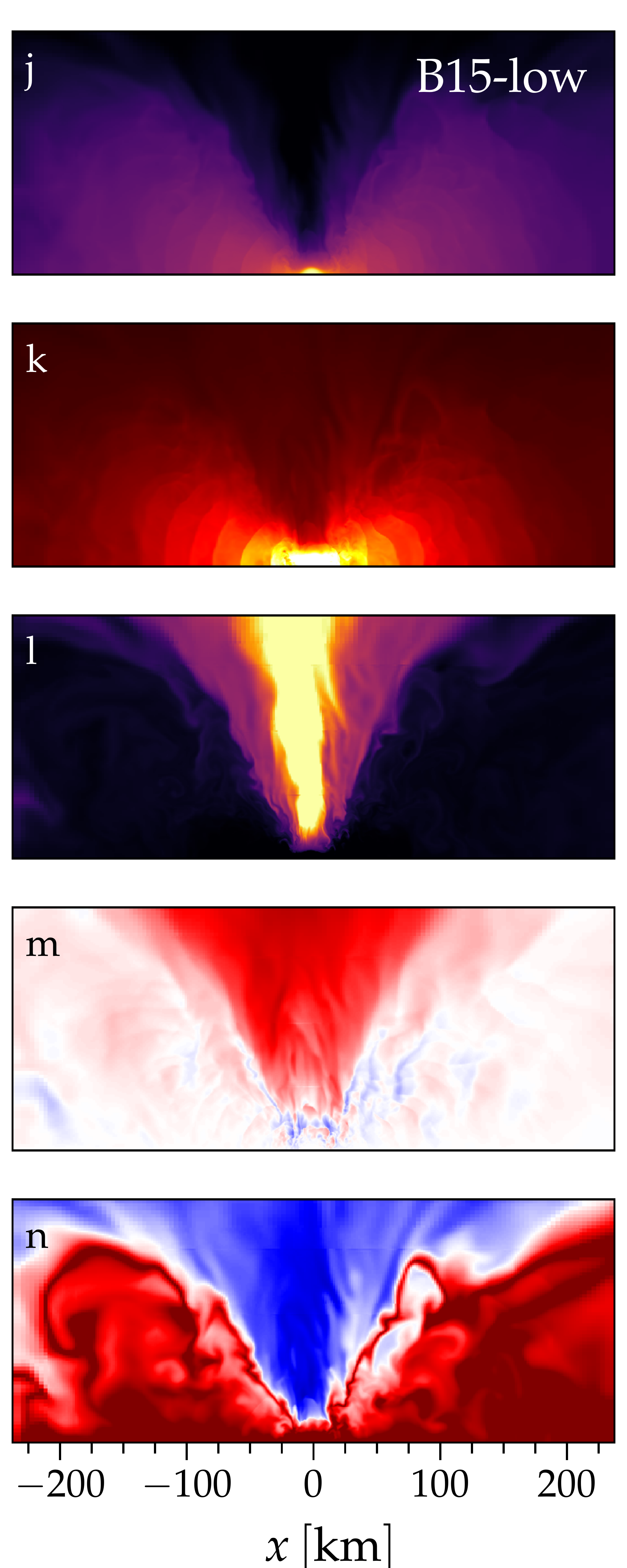} 
	\end{minipage}
	\begin{minipage}{0.268\textwidth}
	\includegraphics[width=\textwidth]{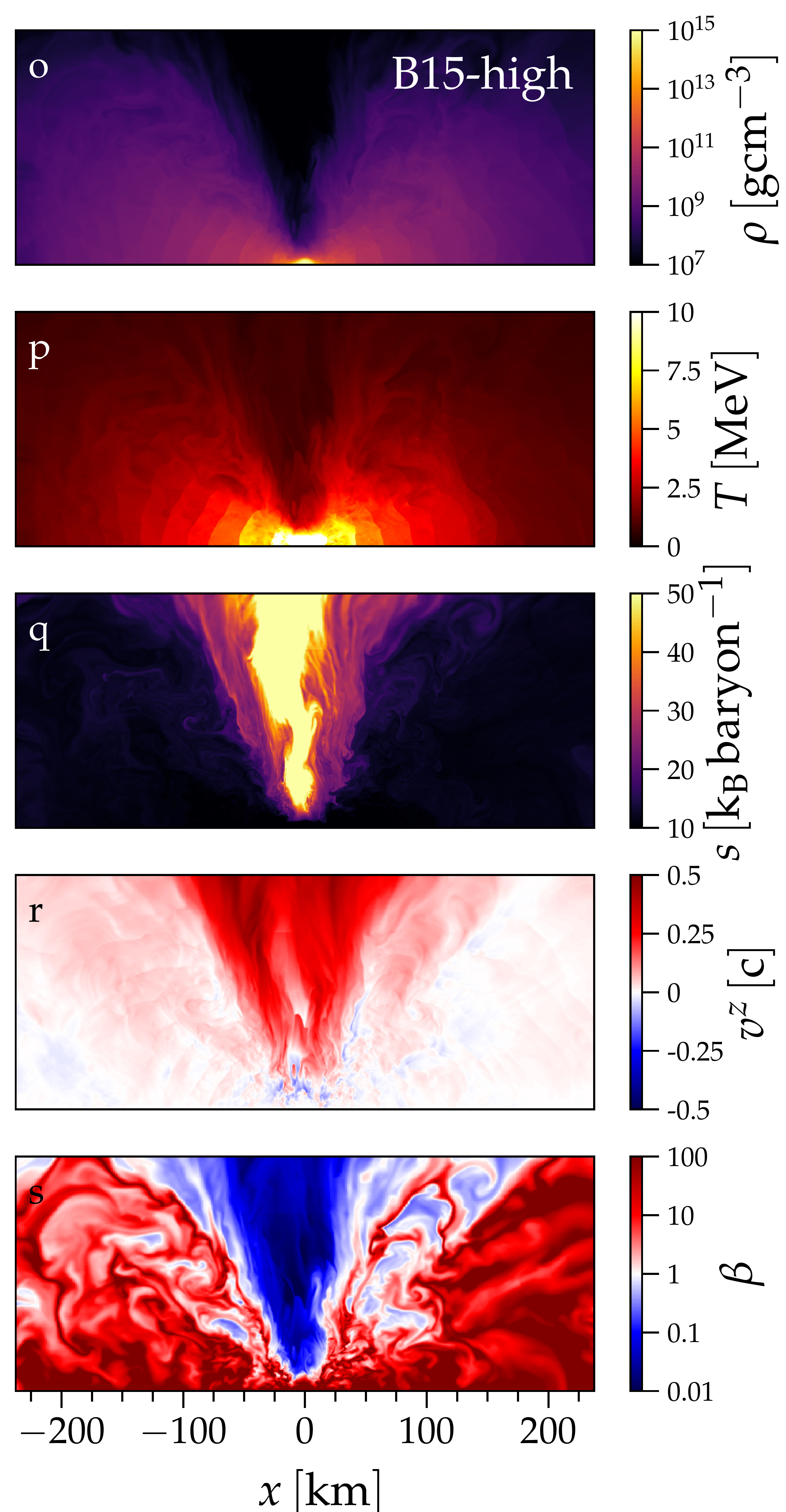} 
	\end{minipage}
\vspace{0.1cm}
	\caption{Meridional slices ($xz$-plane, $z$ being the vertical) of
	density $\rho$, 
	temperature $T$, specific
	entropy $s$, velocity component aligned with rotation axis $v^z$, and
	magnetic pressure $\beta$. Panels a - d show simulation B0, panels e
	- i show simulation B15-nl, panels j-n show simulation B15-low, and
	panels o - s show simulation B15-high (magnetic pressure is only shown
	for simulations B15-nl, B15-low, and B15-high).} \label{fig:2dall} 
\vspace{0.5cm} 
\end{figure*}


\subsection{Outflows}
\label{sec:outflows}

\begin{figure*}[t]
\centering
	\begin{minipage}{0.32\textwidth}
	\includegraphics[width=\textwidth]{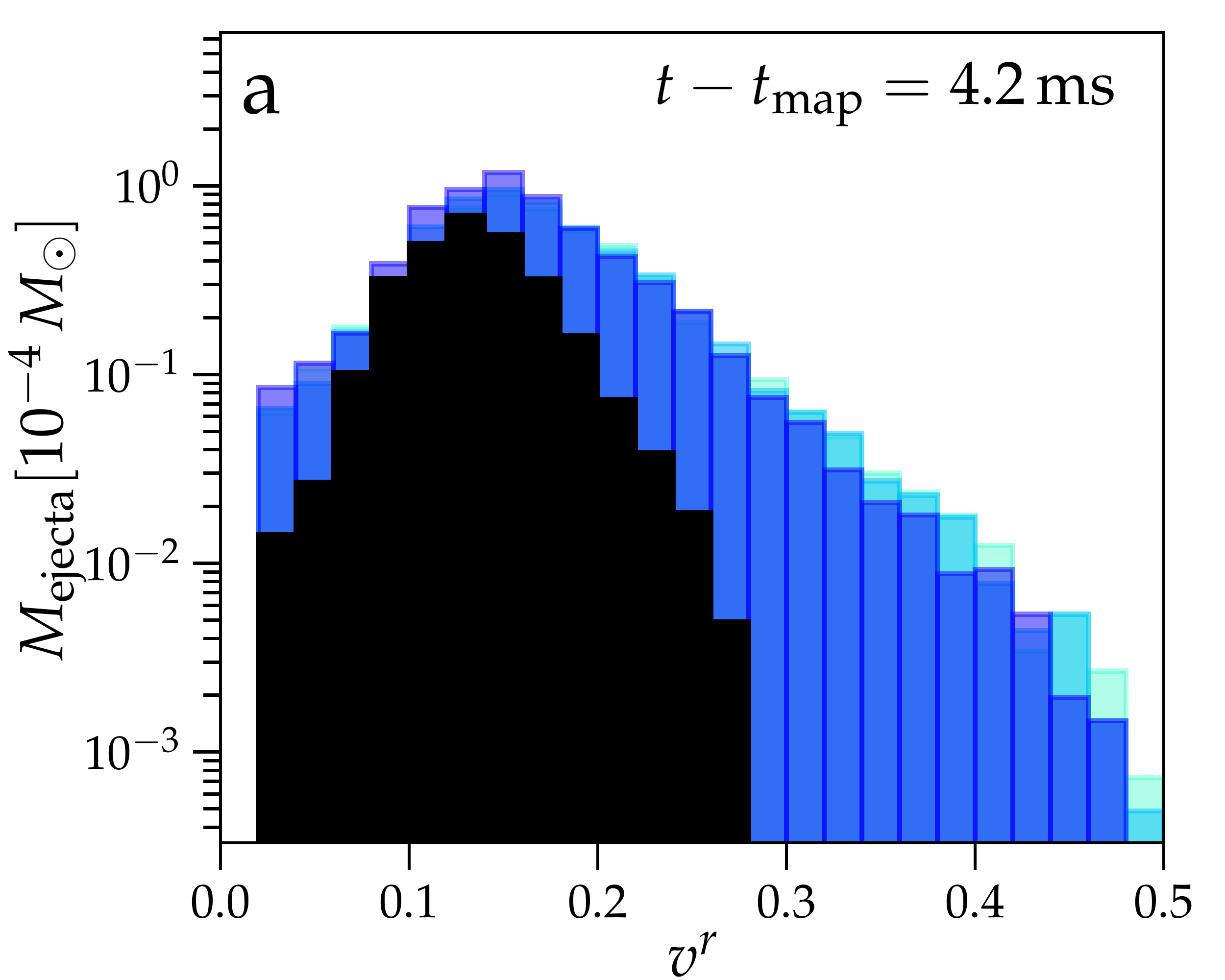} 
	\end{minipage}
	\begin{minipage}{0.32\textwidth}
	\includegraphics[width=\textwidth]{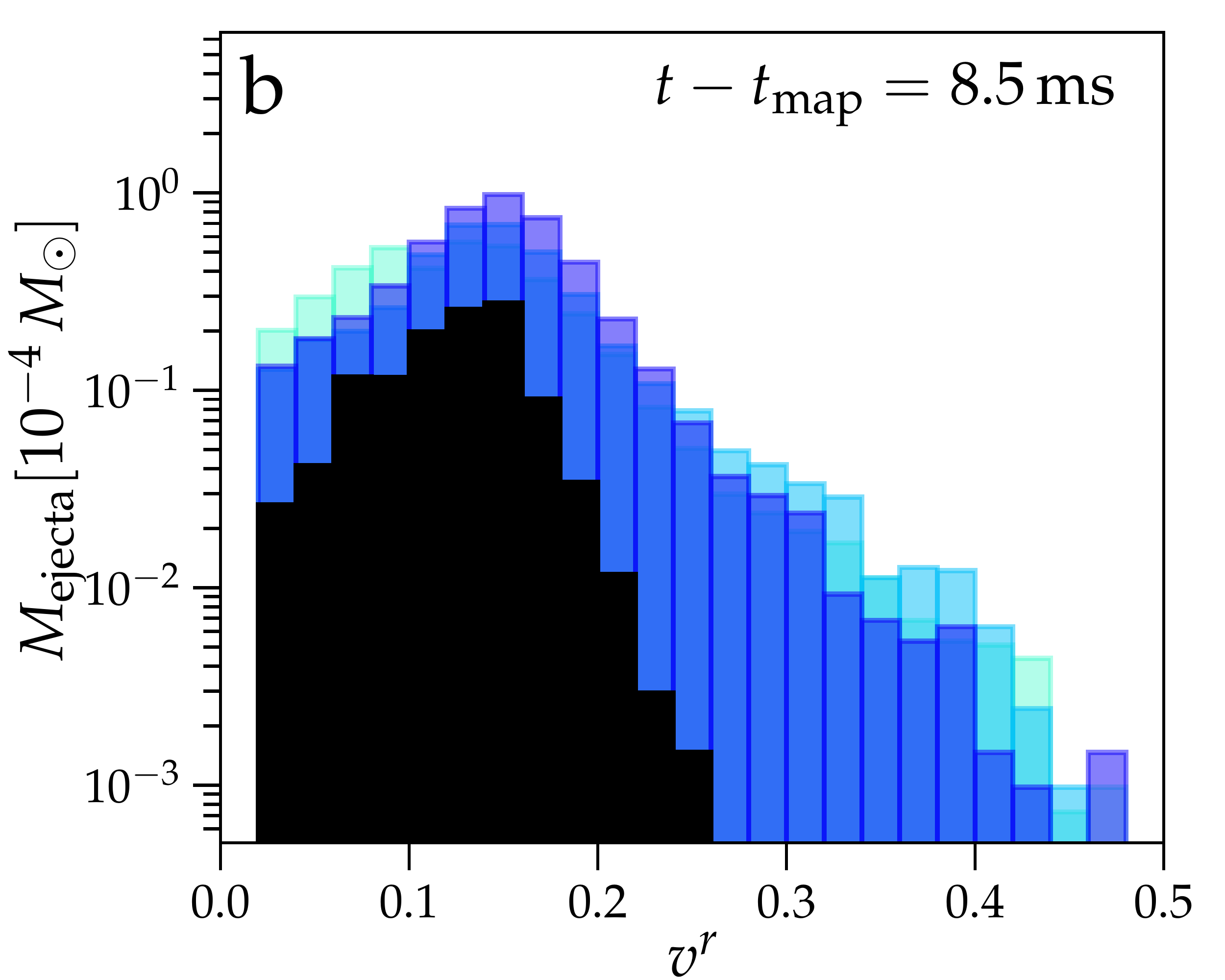} 
	\end{minipage}
	\begin{minipage}{0.32\textwidth}
	\includegraphics[width=\textwidth]{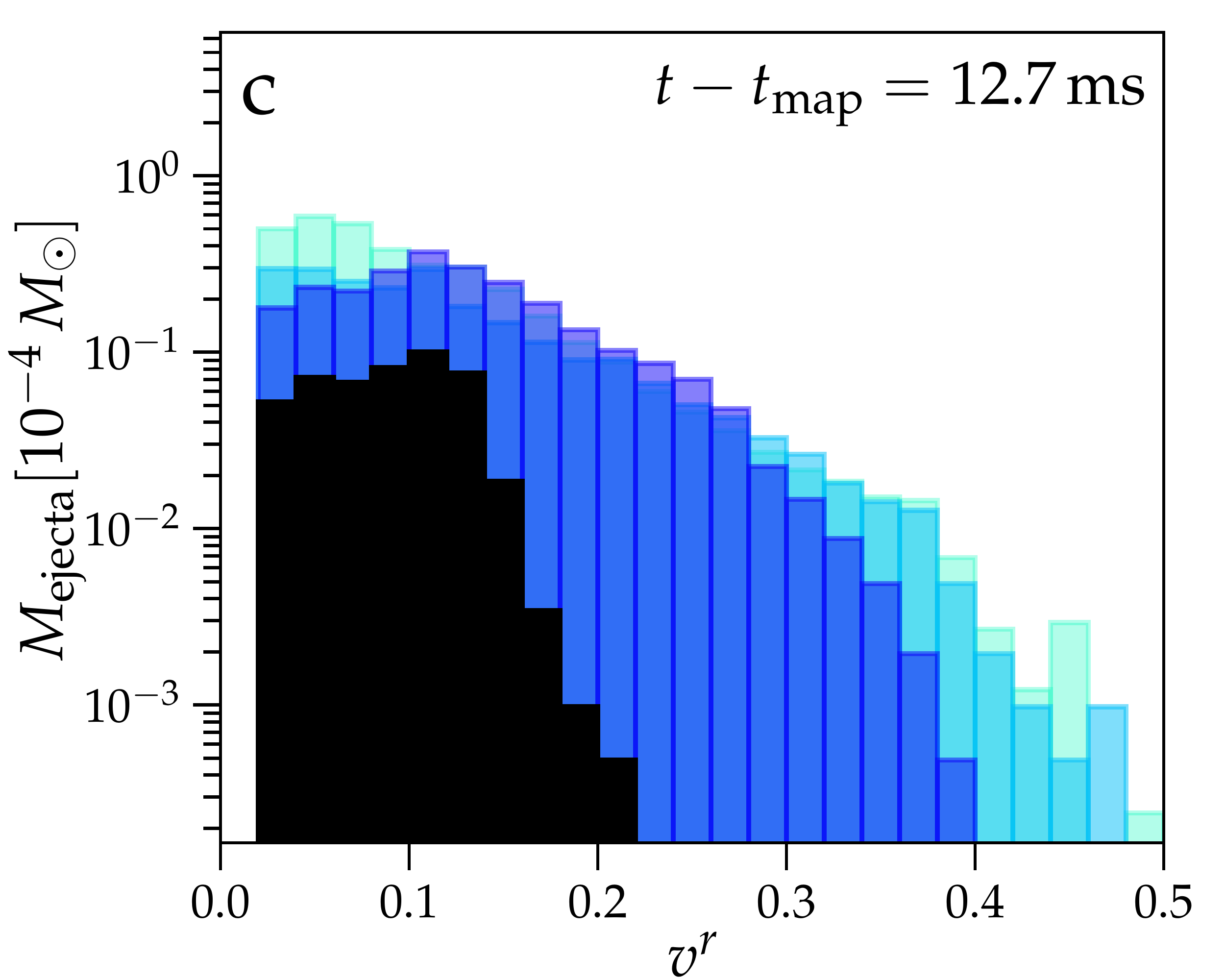} 
	\end{minipage}
	\begin{minipage}{0.32\textwidth}
        \hspace{0.00cm}
	\includegraphics[width=\textwidth]{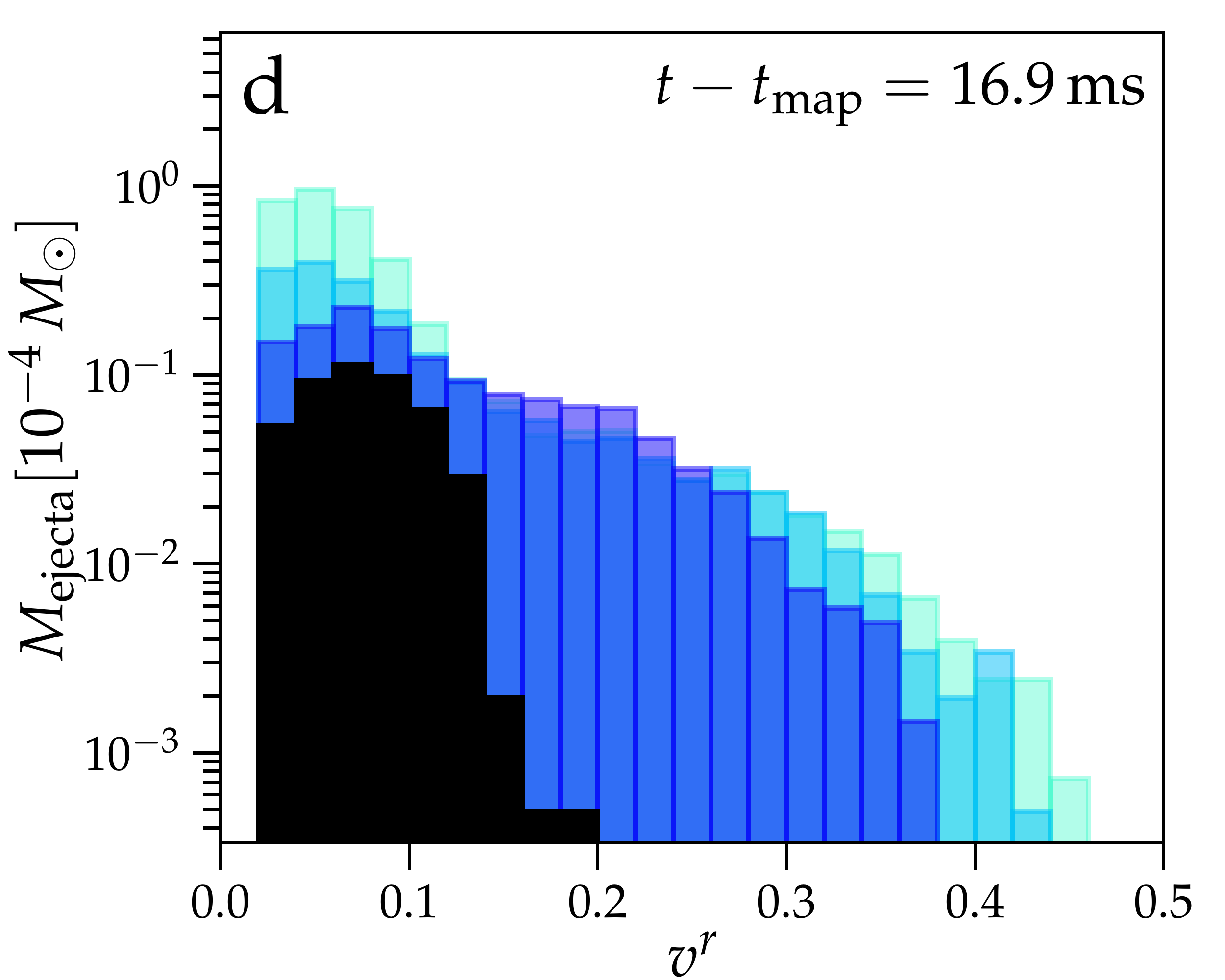}
	\end{minipage}
	\begin{minipage}{0.32\textwidth}
        \hspace{0.00cm}
	\includegraphics[width=\textwidth]{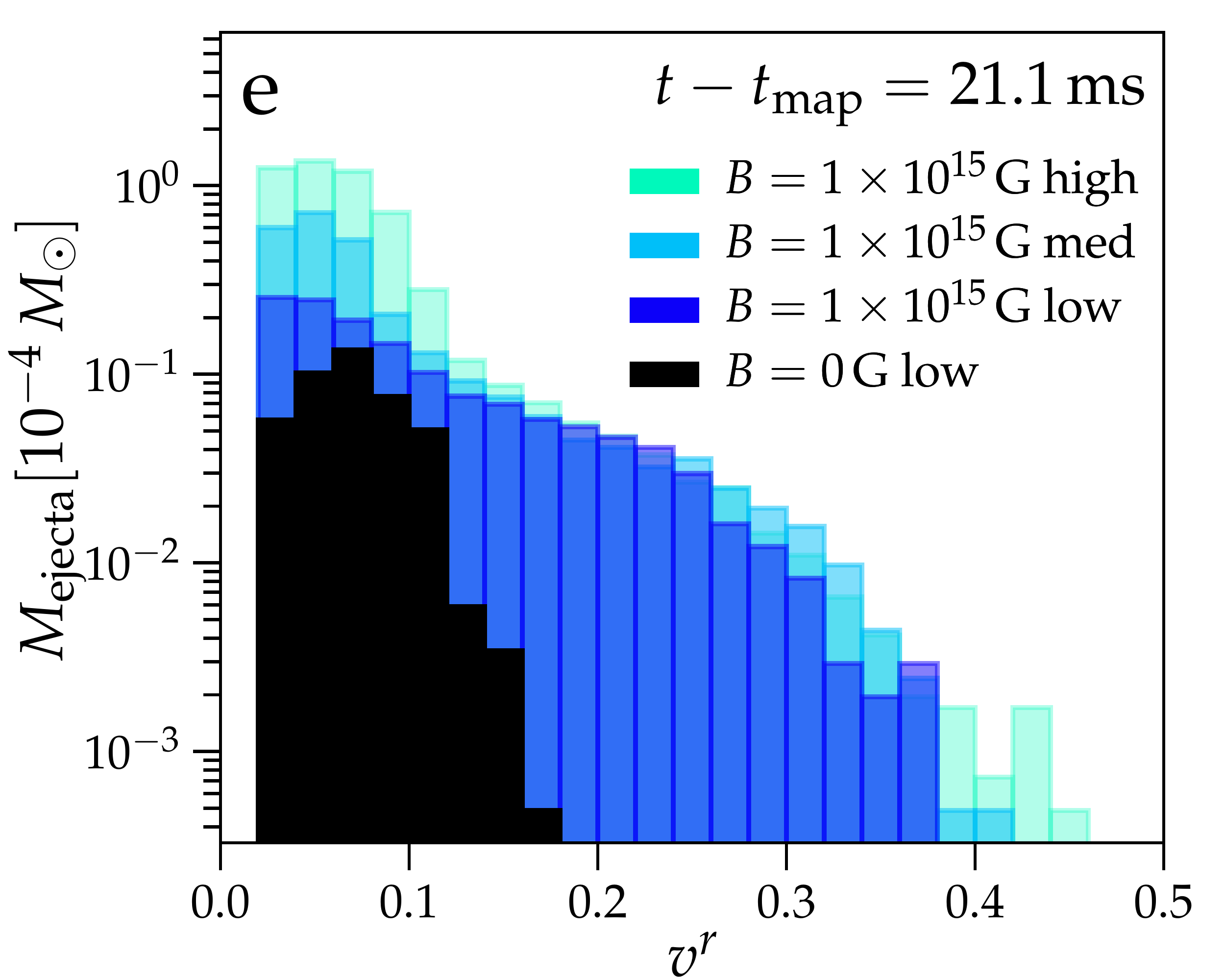} 
	\end{minipage}
	\begin{minipage}{0.32\textwidth}
	\includegraphics[width=\textwidth]{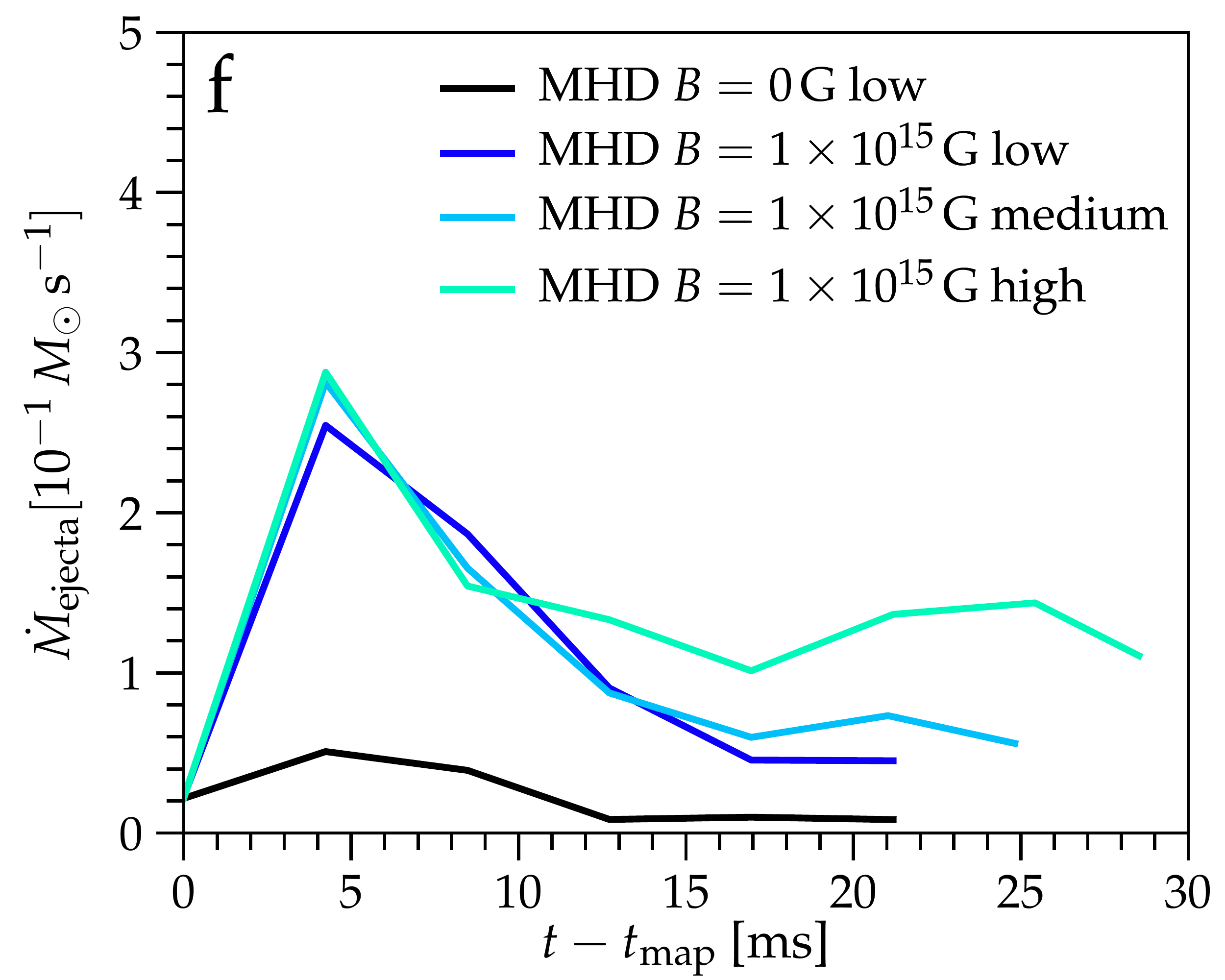}
	\end{minipage}
	\caption{\textbf{Panels a-e:} $v^r$ ($r$ being the radius in spherical
	coordinates) histograms of unbound material at different times during
	simulations B0 (black), B15-low (blue), B15-med (cyan), and B15-high
	(green). We bin 
	the distribution with the mass of the ejected material.
	\textbf{Panel f:} Mass outflow rate $\dot{M}_{\mathrm{ej}}$ as a function of
	post-mapping time $t-t_{\mathrm{map}}$ for simulations B0 (black),
	B15-low (blue), B15-med (cyan), and B15-high (light green). We
	calculate the average (averaged over spheres of $r_0 < r < r_1$)
	outflow rate as $\dot{M}_{\mathrm{ej}} = \int_{r_0}^{r_1} \sqrt{g} \rho
	W v^r dV (r_1-r_0)^{-1}$ with $r_0 = 44.3\, \mathrm{km}$ and $r_1 =
	192.1\, \mathrm{km}$ and only include material in the integral if the
	Bernoulli criterion $-h\, u_t > 1$ indicates that this material is
	unbound.}  \label{fig:vr_hist} 
\vspace{0.5cm} 
\end{figure*}

In Fig.~\ref{fig:2dall} we show renderings of density, temperature,
specific entropy, z-component of velocity and magnetic pressure in 2D
Meridional slices ($xz$-plane, $z$ being the vertical) for simulations B0
(left), B15-nl (center-left), B15-low (center-right), and B15-high (right). We
show the renderings at time $t-t_{\mathrm{map}} \simeq 21.2\, \mathrm{ms}$ for
simulations B0, B15-low, B15-med, and B15-high, and at $t-t_{\mathrm{map}}
\simeq 15.1\, \mathrm{ms}$ for simulation B15-nl to account for the earlier
collapse time in simulation B15-nl (see Fig.~\ref{fig:rhobmax}). There are no large
differences in density structure of the disk when comparing panels a, e, j, and
o. The high-temperature region in the HMNS is more extended
for simulation B15-nl compared to simulations B15-low and B15-high (panels 
f, k, and p).  In all our simulations with neutrino effects the polar region
remains mostly free of baryon pollution. In contrast simulation B15-nl has a
factor 5-10 higher density in the polar region, similarly to the simulations
presented in ~\citep{ciolfi:19,ciolfi:20b}. The HMNS remains more compact in
simulation B15-low and B15-high compared to simulation B15-nl. These
differences are in line with neutrino cooling causing the remnant and its
accretion disk to stay more compact due to reduced thermal pressure. Key
differences between simulation B0 and its magnetized counterparts B15-nl,
B15-low, and B15-high arise in the outflow structure.  While simulation B0
shows an outflow that resembles a high-entropy wind (panel c), simulations
B15-low and B15-high show a collimated, highly magnetized outflow. This is most
clearly visible in panels l and q which depict entropy.  Simulation B15-nl
shows a higher velocity outflow than simulation B0 but lacks a highly
collimated component compared to simulations B15-low and B15-high.  This is
most clearly visible when comparing panels i, n and s which show plasma $\beta
= P / b^2$.  The outflow velocity (panels d,h,m, and r) increases when
comparing simulation B0 ($\sim.0.2c$), B15-nl ($\sim.0.3c$), B15-low ($\sim
0.35c$) and B15-high ($\sim 0.45c$). 

To analyze the properties and composition of the outflows in more detail we
determine unbound material in the simulations via the Bernoulli criterion $-h
u_t > 1$, where $h = (1 + \epsilon + P + \frac{b^2}{2}) / \rho$ is the
relativistic enthalpy of the magnetized fluid. We show histograms of $v^r$ for
the unbound material in Fig.~\ref{fig:vr_hist}. At early times simulations
B15-low, B15-med, and B15-high show a similar distribution in velocity of the
ejecta and significant material at $0.3c < v^r < 0.5c$ (panel a). This is in
contrast to simulation B0 which only shows ejecta with $0 < v^r < 0.28c$. At
later times the velocity distribution of the ejecta shifts slightly for all
simulations. For simulation B15 the highest-velocity component of the ejecta
($v^r > 0.4c$) disappears quickly (panels b - e). Simulation B15-med retains
some of this high-velocity ejecta until later times and simulation B15-high
retains most of the high-velocity ejecta until late time (panels b - e). In
addition all simulations show the appearance of low-velocity material ($v^r <
0.1c$). 

\begin{figure*}[t]
\centering
	\begin{minipage}{0.24\textwidth}
	\includegraphics[width=\textwidth]{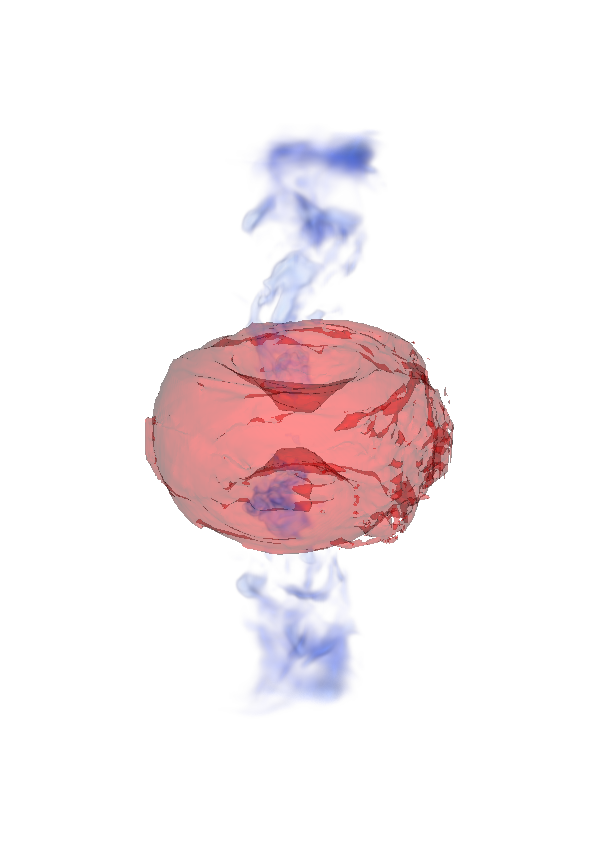}
	\end{minipage}
	\begin{minipage}{0.24\textwidth}
	\includegraphics[width=\textwidth]{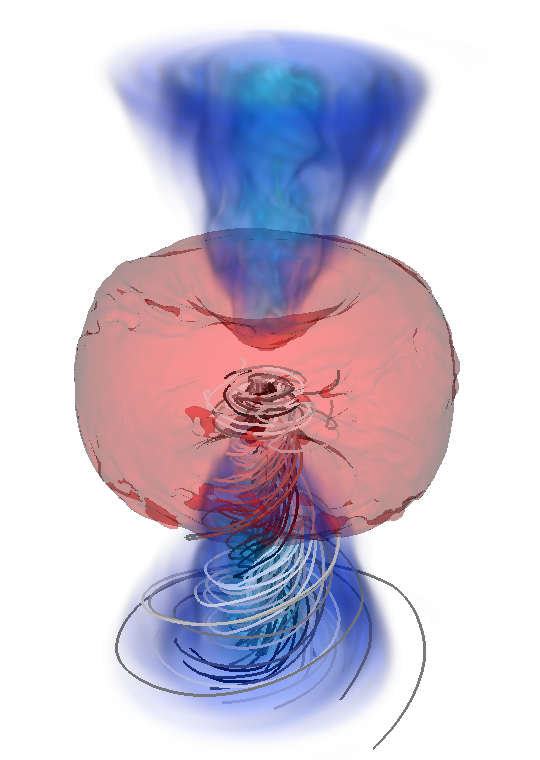}
	\end{minipage}
	\begin{minipage}{0.24\textwidth}
	\includegraphics[width=\textwidth]{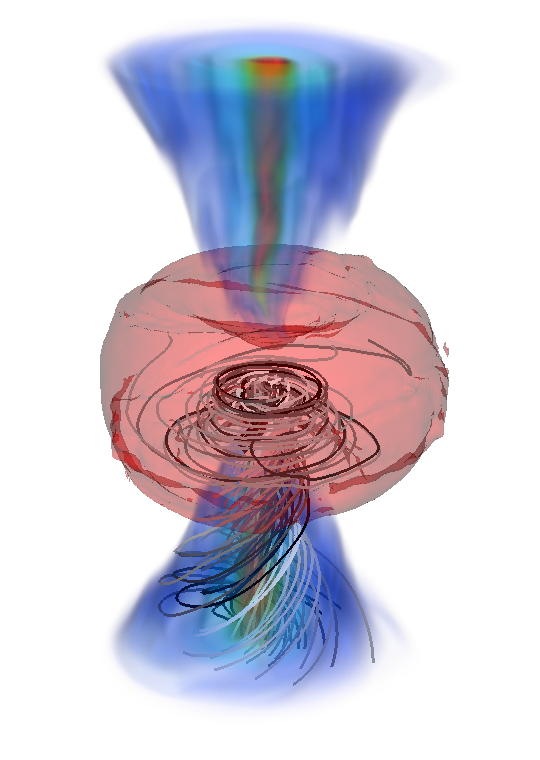}
	\end{minipage}
	\begin{minipage}{0.24\textwidth}
	\includegraphics[width=\textwidth]{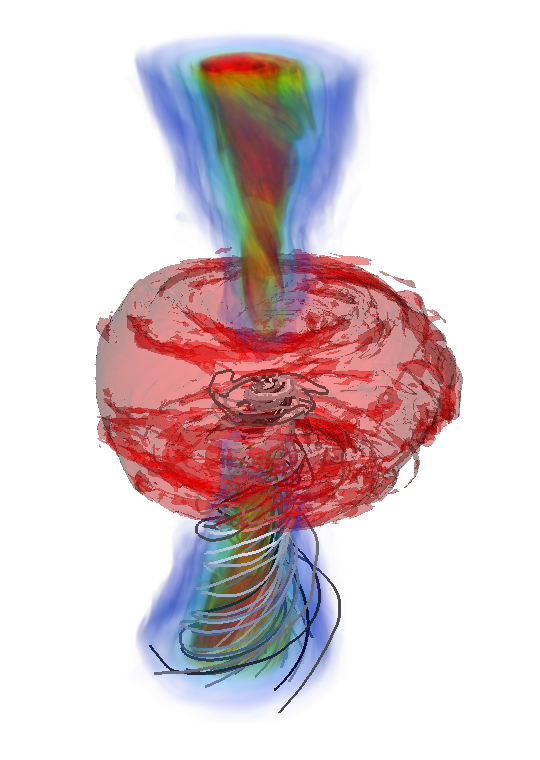} 
	\end{minipage}
\vspace{0.1cm}
	\caption{Volume renderings of the Bernoulli criterion (blue colormap)
	indicating unbound material and the disk contour at $\rho = 10^{10}\,
	\mathrm{g\, cm^{-3}}$ (red) for models B0 (left), B15-nl (center left),
	model B15-low (center right), and B15-high (right). The renderings 
	depict the simulations at $t-t_{\mathrm{map}} = 15.1\, \mathrm{ms}$ for B15-nl, at
	$t-t_{\mathrm{map}} = 19.4\, \mathrm{ms}$ for B0 and B15-low, and at
	$t-t_{\mathrm{map}} = 20.9\, \mathrm{ms}$ for B15-high. The different
	times are chosen to depict the simulations towards the end of
	steady-state operation of the outflows which is at different times $t-t_{\mathrm{map}}$ 
	due to the different collapse times (see Fig.~\ref{fig:rhobmax}).
	Additionally, we show magnetic field lines for simulations B15-nl,
	B15-low, and B15-high in the lower quadrant of the renderings. The
	z-axis is the rotation axis of the HMNS and we show the innermost
	$357\, \mathrm{km}$. The colormap is chosen such that blue corresponds
	to material with lower Lorentz factors $-h\, u_t \simeq 1$, while
	yellow corresponds to material with $-h\, u_t \simeq 1.5$, and red to
	material with $-h\, u_t \simeq 2-5$. We note that for rendering
	purposes we have excluded part of the unbound ejecta in the equatorial
	region.} \label{fig:3d}
\vspace{0.5cm} 
\end{figure*}

To estimate the outflow rate in the simulations we calculate the averaged
mass ejection rate of the outflow $\dot{M}_{\mathrm{ej}} = \int_{r_0}^{r_1}
\sqrt{g} \rho W v^r dV\, (r_1-r_0)^{-1}$ with $r_0 = 44.3\, \mathrm{km}$ and
$r_1 = 192.1\, \mathrm{km}$. We only include material in the integral if the
material is unbound ($-h u_t > 1$). We show $\dot{M}_{\mathrm{ej}}$ as a function
of post-mapping time $t-t_{\mathrm{map}}$ in panel f of Fig.~\ref{fig:vr_hist}.
For all simulations $\dot{M}_{\mathrm{ej}}$ initially rises sharply as the
outflow initially forms before reaching a peak at $t-t_{\mathrm{map}}\simeq 5\,
\mathrm{ms}$. Subsequently $\dot{M}_{\mathrm{ej}}$ evolves towards a
quasi-steady-state that is reached after $t-t_{\mathrm{map}}\simeq 15\,
\mathrm{ms}$. The mass ejection rate for simulation B0 in this phase is
$\dot{M}_{\mathrm{ej}} = 2.4 \times 10^{-3}\,\mathrm{ M_\odot}\ {\rm s}^{-1}$,
which are at the very high end compared to the values predicted by
\citet{thompson:01} for a neutrino-driven wind from the HMNS. For simulations
B15-low we find $\dot{M}_{\mathrm{ej}} = 4.6 \times 10^{-2}\ \mathrm{M_\odot}\
{\rm s}^{-1}$, for simulation B15-med $\dot{M}_{\mathrm{ej}} = 5.6 \times
10^{-2}\ \mathrm{M_\odot}\ {\rm s}^{-1}$, and finally $\dot{M}_{\mathrm{ej}} =
1.2 \times10^{-1}\ \mathrm{M_\odot}\ {\rm s}^{-1}$. These outflow rates are a
factor $\simeq 20$ (for simulations B15-low and B15-med) and a factor $\simeq
100$ (for simulation B15-high) higher than in the hydrodynamic simulation B0
and are consistent with a magnetized wind~\citep{thompson:04b} from the HMNS.

We can also use $\dot{M}_{\mathrm{ej}}$ to estimate the total ejecta amount for
the simulations. For this we average the mass accretion rates over the period
of quasi-steady-state evolution and integrate this over the simulation time.
We find $M_{\mathrm{ej}} = 5.8 \times 10^{-5}\, \mathrm{M_{\odot}}$ for
simulation B0, $M_{\mathrm{ej}} = 1.1 \times 10^{-3}\, \mathrm{M_{\odot}}$ for
B15-low, $M_{\mathrm{ej}} = 1.4 \times 10^{-3}\, \mathrm{M_{\odot}}$ for
B15-med, and $M_{\mathrm{ej}} = 3.5 \times 10^{-3}\, \mathrm{M_{\odot}}$ for
B15-high. These ejecta masses make the ejecta from the HMNS important when
compared to the dynamical ejecta $10^{-4}\, \mathrm{M_{\odot}} <
M_{\mathrm{ej}} < 10^{-2}\, \mathrm{M_{\odot}}$ and winds driven from a BH
accretion disk.

To illustrate the nature and geometry of the outflow, accretion disk, and
magnetic field structure we show 3D volume renderings of the Bernoulli
criterion in combination with an isocontour plot for a density of $10^{10}\,
\mathrm{g\, cm^{-3}}$ and streamlines of the magnetic field for simulations
B0, B15-nl, B15-low, and B15-high in Fig.~\ref{fig:3d}. These renderings make
the additional emergence of a mildly relativistic jet in simulation B15-low and
B15-high immediately obvious (narrow red funnel aligned with rotation axis
(z-axis)). This is in contrast to simulation B15-nl. The jet in simulation
B15-low reaches a maximum Lorentz factor $\simeq 2$ while the jet in simulation
B15-high reaches a Lorentz factor $\simeq 5$. We also calculate the average
luminosity of the jet as $L_{\mathrm{ejecta}} = \int_{r0}^{r1} T^{0i} r_i dV$
where we include only material in the integral that has $-h u_t > 2$. During
steady-state operation we find $L_{\mathrm{ejecta}} \sim 10^{50}\, \mathrm{erg
\, s^{-1}}$ for simulation B15-low and $L_{\mathrm{ejecta}} \sim 10^{51}\, 
\mathrm{erg\, s^{-1}}$ for simulation B15-high, while simulation B15-nl does not
have material with $-h u_t > 2$. These results indicate that neutrino effects,
i.e. neutrino cooling reducing baryon pollution in the polar region, are
important for the emergence of the jet and that turbulent magnetic field
amplification can significantly boost its Lorentz factor and energetics.

\section{Discussion} 
\label{sec:discussion}

We have carried out dynamical GRMHD simulations of a magnetized hypermassive NS
formed in a BNS merger including a nuclear EOS and neutrino cooling and
heating.  We have run simulations at three different resolutions of up to $h =
55\, \mathrm{m}$ and reference simulations with no magnetic field and no
neutrino physics. The highest-resolution simulation is designed to fully
resolve magnetoturbulence driven by the MRI. We have run all the simulations to
collapse to a BH.

We find an outflow that is consistent with a magnetized
wind~\citep{thompson:04b} from the HMNS that ejects neutron-rich
material along the rotation axis of the remnant with an outflow rate
$\dot{M}_{\mathrm{ej}} \simeq 1 \times 10^{-1}\, \mathrm{M_{\odot}\,
\mathrm{s^{-1}}}$. This leads to a total ejecta mass of $3.5 \times 10^{-3}\,
\mathrm{M_{\odot}}$ for the binary configuration we have studied in this paper.
We can also use the average outflow rate calculated during quasi-steady state
operation to estimate the ejecta mass for binary configurations that leave
behind HMNSs that collapse at later times. For longer-lived remnants the total
ejecta mass can therefore be the dominant ejecta component when compared to the
dynamical ejecta $10^{-4}\, \mathrm{M_{\odot}} < M_{\mathrm{ej}} < 10^{-2}\,
\mathrm{M_{\odot}}$ and winds driven from a BH accretion disk.  

The broad distribution in velocity space of the ejecta with a significant
fraction of material with velocities in the range of $0.3c < v^{r} < 0.5c$ sets
it apart from the dynamical ejecta $v^r < 0.3c$ and winds driven from an
accretion disk $v^r < 0.1c$~\citep{fahlman:18}. Thus magnetized winds, possibly
in combination with spiral-wave driven outflows \citep{nedora:19}, can
explain the blue component of the kilonova in GW170817, as
anticipated by \citet{metzger:18a}. Taking into account the outflow rates
observed in the simulations, results from other published numerical studies
\citep{shibata:17a, radice:18a, nedora:19}, and the inferred overall mass
ejected by the NSM in GW170817, our results suggest a plausible scenario
in which the merger remnant collapsed to BH on a timescale of $O(100\ {\rm
ms})$. This is consistent with earlier interpretation of the event based on
both the red and blue kilonova observations~\citep{margalit:17}.

The magnetic field enables the launch of a jet in all simulations with
neutrino effects. The emergence of this jet is aided by neutrino cooling which 
reduces baryon pollution in the polar region. We also find that MRI-driven
turbulence is effective at amplifying the magnetic field in the shear layer
outside of the HMNS core to $10^{16}\, \mathrm{G}$ and that this ultra-strong
toroidal field can significantly boost the Lorentz factor of the jet. In our
highest-resolution simulation the jet reaches a terminal Lorentz factor of
$\simeq 5$, is mildly relativistic, and the corresponding luminosity is $\simeq 10^{51}\,
\mathrm{erg\, s^{-1}}$. The Lorentz factor measured from our simulations is
only a conservative lower estimate as we did not include full neutrino
transport. Neutrino pair-annihilation may lead to ejected material being less
baryon-rich than in our simulations~\citep{fujibayashi:17} and this can boost
the Lorentz factor to the relativistic sGRB regime~\citep{just:16}.  With this
in mind our simulations indicate that magnetars formed in NS mergers are a
promising sGRB engine.

\section*{Acknowledgments}

The authors would like to thank M.~Campanelli, F.~Foucart, J.~Guilet,
E.~Huerta, D.~Kasen, S.~Noble, and E.~Quataert, and A.~Tchekhovskoy for
discussions and support of this project. The authors would like to thank the
anonymous referees for useful suggestions improving the manuscript. PM
acknowledges support by NASA through Einstein Fellowship grant PF5-160140. SB
acknowledges support by the EU H2020 under ERC Starting Grant, no.
BinGraSp-714626. The simulations were carried out on NCSA's BlueWaters under
NSF awards PRAC OAC-1811352 (allocation PRAC\_bayq), NSF AST-1516150
(allocation PRAC\_bayh), and allocation ILL\_baws, and TACC's Frontera under
allocation DD FTA-Moesta. 
Figures were prepared using \texttt{matplotlib}~\citep{Hunter:2007}
and \texttt{VisIt}~\citep{HPV:VisIt}. Research at Perimeter Institute is supported
in part by the Government of Canada through the Department of Innovation,
Science and Economic Development Canada and by the Province of Ontario through
the Ministry of Colleges and Universities. 


\bibliographystyle{aasjournal}



\end{document}